\documentclass[aps,prl,reprint,superscriptaddress,nofootinbib]{revtex4-1}

\usepackage{graphicx}
\usepackage{dcolumn}
\usepackage{bm}
\usepackage{amssymb}
\usepackage{verbatim}
\usepackage{epstopdf}

\newcommand{\beq}{\begin{equation}}
\newcommand{\eeq}{\end{equation}}

\newcommand{\bra}{\langle}
\newcommand{\ket}{\rangle}

\begin{document}

\title[Measuring and engineering entropy and spin squeezing]{Measuring and engineering entropy and spin squeezing in weakly linked Bose-Einstein condensates}

\author{F.~Cattani}
\affiliation{School of Mathematics, University of Southampton, SO17
  1BJ, Southampton, UK}
\author{C.~Gross}
\affiliation{Max-Planck-Institut f\"{u}r Quantenoptik, 85748 Garching, Germany}
\email{christian.gross@mpq.mpg.de}
\author{M.~K.~Oberthaler}
\affiliation{Kirchoff-Institut f\"{u}r Physik, Universit\"{a}t
  Heidelberg, Im Neunheimer Feld 227, 69120 Heidelberg, Germany}
\author{J.~Ruostekoski}
\affiliation{School of Mathematics, University of Southampton, SO17 1BJ, Southampton, UK}
\date{\today}

\begin{abstract}
We propose a method to infer the
single-particle entropy of bosonic
atoms in an optical lattice and to study the local evolution of
entropy, spin squeezing, and entropic inequalities for entanglement
detection in such systems.
This method is based on experimentally feasible measurements of
  non-nearest-neighbour coherences.
We study a specific example of dynamically controlling atom tunneling
between selected sites and show that this could potentially also
improve the metrologically relevant spin squeezing.
\end{abstract}


\maketitle

The quest for novel cooling schemes to control the entropy of ultracold
atoms in optical lattices is attracting considerable interest because
thermal and quantum fluctuations limit the use of these systems for
quantum simulation or quantum metrology.
For example, the
experimental observation of magnetic ordering, a milestone for the quantum
simulation of spin systems, is  hindered by finite entropy in the
system~\cite{ref:werner, ref:joerdens, ref:entropyBosons, ref:greif}.
In metrology, finite entropy limits the amount of
achievable spin squeezing, a useful resource for quantum-enhanced
high precision measurements~\cite{wineland, bouyer, holland,giovannetti,ref:gross}.
Not only controlling the entropy is challenging; it is also difficult to
measure on a microscopic level.
Only in the low
atomic filling regime of the Mott-insulating phase a mapping of observable
on-site atom number fluctuations to single-particle entropy has been achieved~\cite{ref:sherson2010},
stimulating interest in entanglement detection by entropy measurements~\cite{ref:abanin,ref:zoller}.
Recent experiments show that individual atoms can now be manipulated
on a single-spin level at individual lattice sites~\cite{ref:weitenberg} and, in principle,
the entropy can be locally engineered in order to effectively cool the system~\cite{ref:bernier}.

 \begin{figure}[t]
 \begin{center}
 \includegraphics[width=0.8\columnwidth]{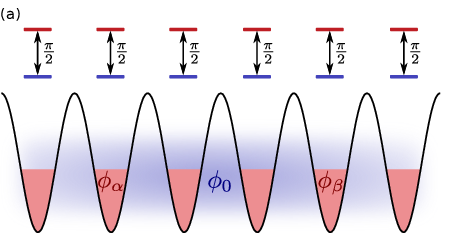}
 \includegraphics[width=0.8\columnwidth]{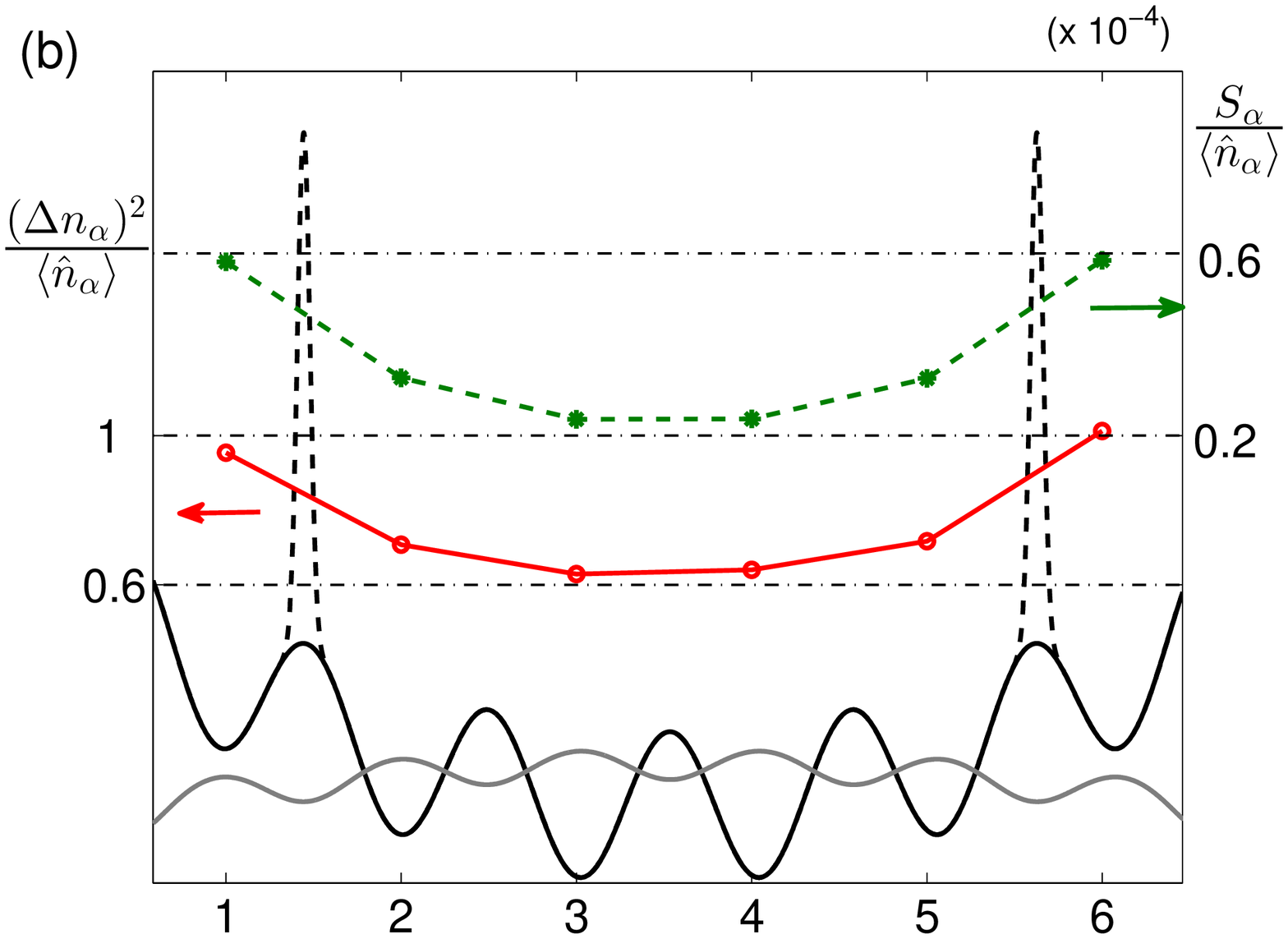}
 \end{center}
 \caption{\label{fig:1}
(a) A schematic illustration of the relative
   phase measurement scheme. (b) Entropy engineering in the optical
   lattice where the tunneling of
   the atoms from the outermost sites to
   the rest of the system is suppressed, e.g., by applying symmetrically-positioned,
   tightly-focused, blue-detuned laser beams.
   We show the
   combined  lattice potential and the optical barriers
   (black solid and dashed lines), the initial atom density before the sites are
   decoupled (solid grey line), the single-particle entropy
   $S_\alpha/\langle \hat{n}_\alpha\rangle$
   per atom in  site $\alpha$ (green dashed line with stars), and
   the normalized on-site atom number fluctuations
     $\left(\Delta n_\alpha\right)^2/\langle \hat{n}_\alpha\rangle$
   (red solid line with circles) in the initial state for each lattice
   site. The temperature is assumed to be $4$ nK.}
 \end{figure}

We propose techniques for accurate local detection and control of a single-particle entropy
of a bosonic atomic gas in an optical lattice in the large filling regime, opening up novel avenues
for estimating entropy-based quantities and engineering entropy.
In the case of non-negligible tunneling of atoms between adjacent sites, access to the inter-site
coherence is crucial for the determination of the entropy. An
essential ingredient of our proposed entropy measurement is a
technique for a spatially resolved measurement of long-range coherence
of the lattice system based on a matter-wave
homodyne measurement with respect to a reference condensate. Using
this method we theoretically show how the entropy and entropic inequalities of entanglement
detection can be
accurately estimated in experimentally realistic cases \cite{ref:gross, ref:gross_pra} by the atom number
in individual sites and the relative phase coherence between the atoms in different sites.
The proposed coherence measurement may also lead to improved detection of spin squeezing of atoms between
different sites, with potential applications to high-precision measurements.
It also allows the detection of spin squeezing between {\em non-nearest-neighbour} sites and other spatially separated regions.
Moreover, we study a specific example of controlling entropy by locally tailoring the trapping
potential.
We show how the coupling between the high-entropy regions and the rest
of the system is adjusted by suddenly applying laser barriers between
the central
and outermost sites, altering the entropy distribution and affecting
spin squeezing in the system. An illustration of the proposed  scheme
is shown in figure~\ref{fig:1} together with
the variation of the entropy
and the on-site atom number fluctuations in the initial state:
The barriers suppress the interactions between the atoms in the
central sites and the outer-well atoms that exhibit stronger thermal
and quantum fluctuations. We present example simulations in which
injecting
the laser barriers leads to reduced spin fluctuations and improved
spin squeezing. We also find spin squeezing between atoms
occupying spatially distant sites.\\
We consider Bose-condensed atoms confined in an elongated trap in
which we neglect density fluctuations in the radial direction. In the
axial direction the atoms experience a combined harmonic and
optical lattice potential of a few sites, forming an array of weakly linked
Bose-Einstein condensates (BECs). As the lattice height is increased
the atom number fluctuations in each site are reduced and the system
can exhibit metrologically relevant spin squeezing between the atoms
in adjacent sites \cite{ref:gross}.  Each lattice site occupies a multi-mode BEC and the
atom number statistics in such a system is influenced by interactions
between several modes in each site~\cite{ref:gross_pra} (in the simulations we include nine
vibrational modes per site). By
numerically solving for the amplitudes of the wavefunctions
$|\varphi_{\alpha j}\rangle$, determined by the lattice sites $\alpha$ and
vibrational levels $j$, we can construct the atom number fluctuations,
phase coherence, and the entire density matrix of the system,
\begin{equation}
\hat\rho=\sum_{\alpha j \beta l}p_{\alpha j,\beta l} |\varphi_{\alpha,j} \rangle\langle\varphi_{\beta,l}|\,.
\end{equation}
This can then be used to evaluate the single-particle
von Neumann entropy
\begin{equation}
S=-\textrm{Tr}(\hat{\rho}\log \hat{\rho})\,.
\label{eq:entropy}
\end{equation}
The density matrix elements $p_{\alpha j,\beta l} $ are determined by the mode populations and their
relative phase coherence (see~\ref{sec:twa}).

Experimentally, it is challenging to measure
populations of the individual vibrational levels or the relative phase
coherence between them. 
In order to circumvent the need
to gather such detailed information, we will show that in experimentally
realistic situations the single-particle entropy may be estimated by the
atom number and the relative phase coherence between the atoms in
different sites that are obtained by averaging over the vibrational level
structure in each site.

In the experiments~\cite{ref:gross,ref:gross_pra} a high-precision
optical absorption imaging provided site-resolved detection by
integration of the imaged atom density. Moreover, local interference
measurements were performed after a short condensate expansion time
allowing only the atoms from adjacent sites to overlap. The relative
phase coherence of the atoms between the adjacent sites was then
inferred from the phase variance of the interference pattern. In order
to detect the long-range phase coherence in the lattice and extract
sufficiently accurate information of the entropy, we propose a
{\em matter-wave homodyne} measurement scheme for the atoms:
We consider a system where the bosonic atoms in the lattice are
surrounded by a BEC in a different internal
state, e.g., a different hyperfine level for alkali-metal atoms or a
metastable triplet state for alkaline-earth or rare-earth metal atoms (figure~\ref{fig:1}).
The atoms in this second internal state are
assumed to experience a much weaker lattice potential or weaker interactions, so that
their phase coherence is well preserved over the entire lattice length.
Experimentally, inter-species interaction might be controlled by a Feshbach resonance
or by adjusting the spatial overlap between the two species, e.g., with superlattices~\cite{ref:superlattice1,
  ref:superlattice2}.
The surrounding BEC serves as a common phase reference in analogy
to local oscillators in quantum optical homodyne measurements.
The phase coherence of the atoms in each lattice site can be locally
determined by interfering the atoms with the reference condensate,
e.g., by Raman transitions between the
hyperfine levels~\cite{ref:twinatoms}.
Due to negligible phase fluctuations in the reference condensate,
local interference measurements of the lattice atoms with the
reference condensate provide information about the
relative phase fluctuations between the atoms in distant
sites.
The proposed scheme has the advantage of precise coherence readout by particle counting
as experimentally problematic shot-to-shot fluctuations of the
magnetic field (which lead to excess phase fluctuations) do not
disturb such measurements as long as they are spatially
homogeneous. 
The measurement method could also be suitable for
2D lattice systems when combined with recently developed high
resolution imaging techniques~\cite{ref:sherson2010, ref:greiner2009}.

In order to demonstrate how the long-range coherence and atom number detection can be used to infer
locally a single-particle entropy and spin squeezing in an
experimentally realistic system we study a specific example of
dynamically adjusting the atom tunneling between central and outermost
sites (see figure~\ref{fig:1}).
We assume that the atoms are initially confined in a thermal equilibrium state in a shallow
lattice. We then simulate the resulting dynamics when the coupling of the outermost
sites to the rest of the system is suppressed by a rapid injection of
laser barriers, followed by a slow ramp up of the lattice potential. The
increase in the lattice depth results in reduced atom number fluctuations and
stronger spin squeezing between atoms in adjacent sites. The laser barriers alter the entropy distribution
in the system, as the initial thermal fluctuations in the trap are not uniform (figure~\ref{fig:1}).
Our numerical simulations are based on the truncated Wigner approximation
(TWA) \cite{DRU93,Steel,SIN02,ref:isella06,ref:review,polkovnikov,martin10}, using
an approach similar to the one introduced in Ref.~\cite{ref:gross_pra}. Here thermal and quantum
fluctuations of the atoms in the stochastic initial state are calculated by self-consistently solving the
ground-state and excited-state populations within the Hartree-Fock-Bogoliubov
approximation \cite{ref:hfb1,ref:HFB}. During the time dynamics the field amplitudes
in the wavefunction basis $|\varphi_{\alpha j}\rangle$ are obtained by projecting from the numerically calculated
stochastic Wigner field (see~\ref{sec:twa}).

We take the experimental parameters that were used to observe
spin squeezing between the atoms in adjacent sites \cite{ref:gross} in which case $N=5300$ atoms
were confined in a combined 1D lattice potential, with the spacing
$d\simeq 5.7\mu$m, and an elongated harmonic trap with the frequencies
$\omega \simeq 2\pi \times 21 $ Hz and $ \omega_\perp \simeq 2\pi
\times 427$ Hz ($\omega \ll\omega_\perp$), so that about 95\% of the
atoms occupied
the six central sites.
In Ref.~\cite{ref:gross_pra} the 1D TWA model provided a good
qualitative agreement with the experimental findings of the on-site
and the relative atom number fluctuations. Here we use the same approach with the potential
\beq
V(x) = \frac{m}{2} \omega^2 x^2 + s E_R\cos^2\left({\pi x\over d}\right)\,,
\eeq
The initial lattice height of $24\,E_R$
(with the recoil energy $E_R=\hbar^2\pi^2/2md^2$) is slowly turned up to $72\,E_R$.
We study two different ramping speeds $15.6$ Hz/ms and $17.2$ Hz/ms.
The strength of the nonlinear atom-atom interaction is
given by $g_{1D} N = 487 \hbar \omega l$, where
$g_{1D}=2\hbar \omega_\perp a$, $a$ is the $s$-wave scattering length, and $l=(\hbar/m\omega)^{1/2}$.
We consider the selection of the four central sites by injecting narrow, blue-detuned
laser beams, or a {\em scissors} potential, before the ramping up of the lattice.
The laser potential can
be modeled by two symmetric Gaussian intensity distributions centered at $\pm x_b$
\beq
V_b = s_b E_R \left[ \exp \left(-{(x-x_b)^2\over 2 d_b^2}\right) +\exp \left(-{(x+x_b)^2\over 2 d_b^2}\right)\right] \,,
\eeq
with the waist $2d_b=600$nm ($1/e^2$ intensity radius) that can be achieved by
diffraction-limited focusing of a laser with wavelength in the visible
light region. We vary the barrier
height $s_b$ and the cutting is fast
compared to any other time scale so that the system has no time to
relax during the process.

 \begin{figure}[t]
 \begin{center}
\includegraphics[width=0.45\columnwidth]{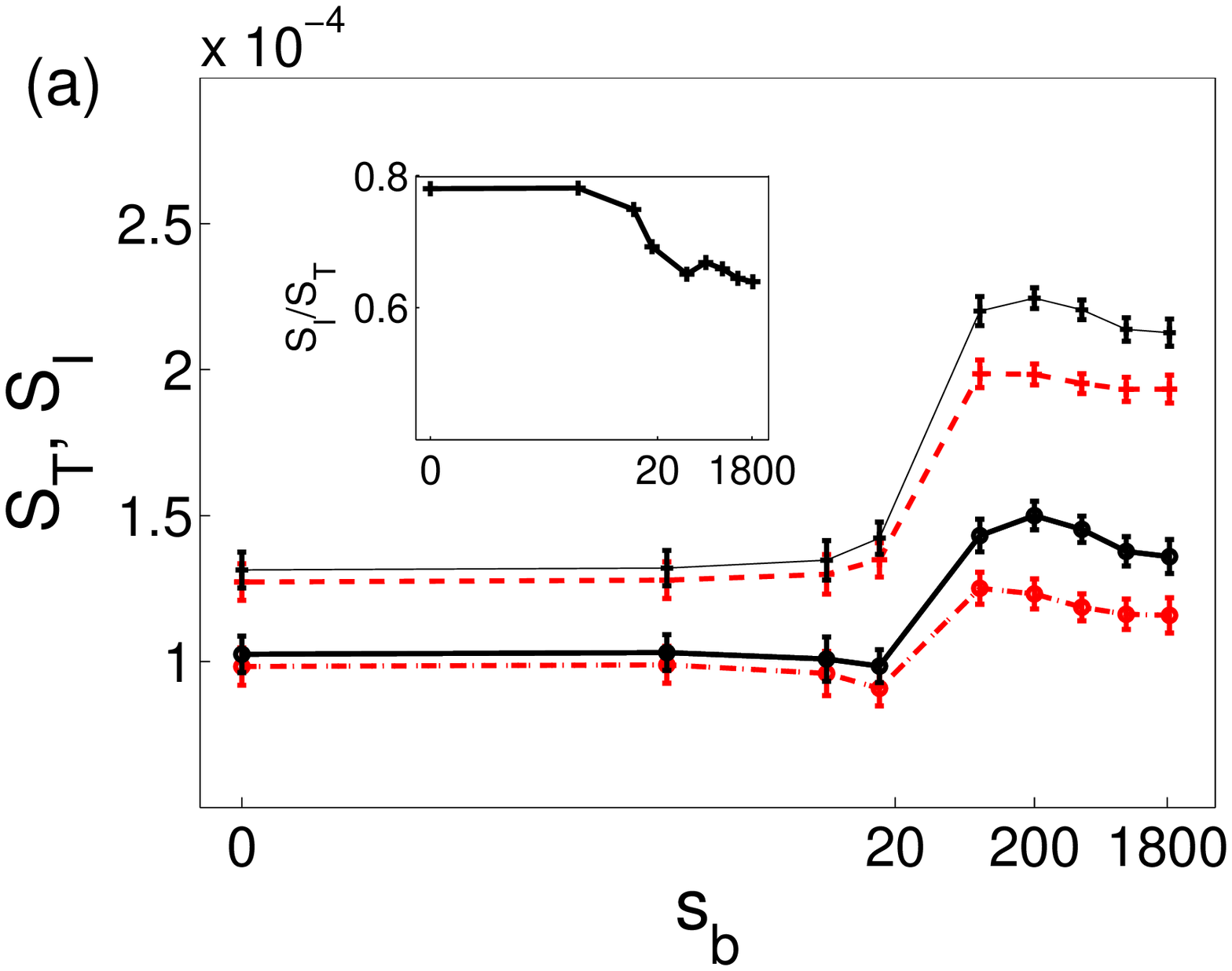}\includegraphics[width=0.45\columnwidth]{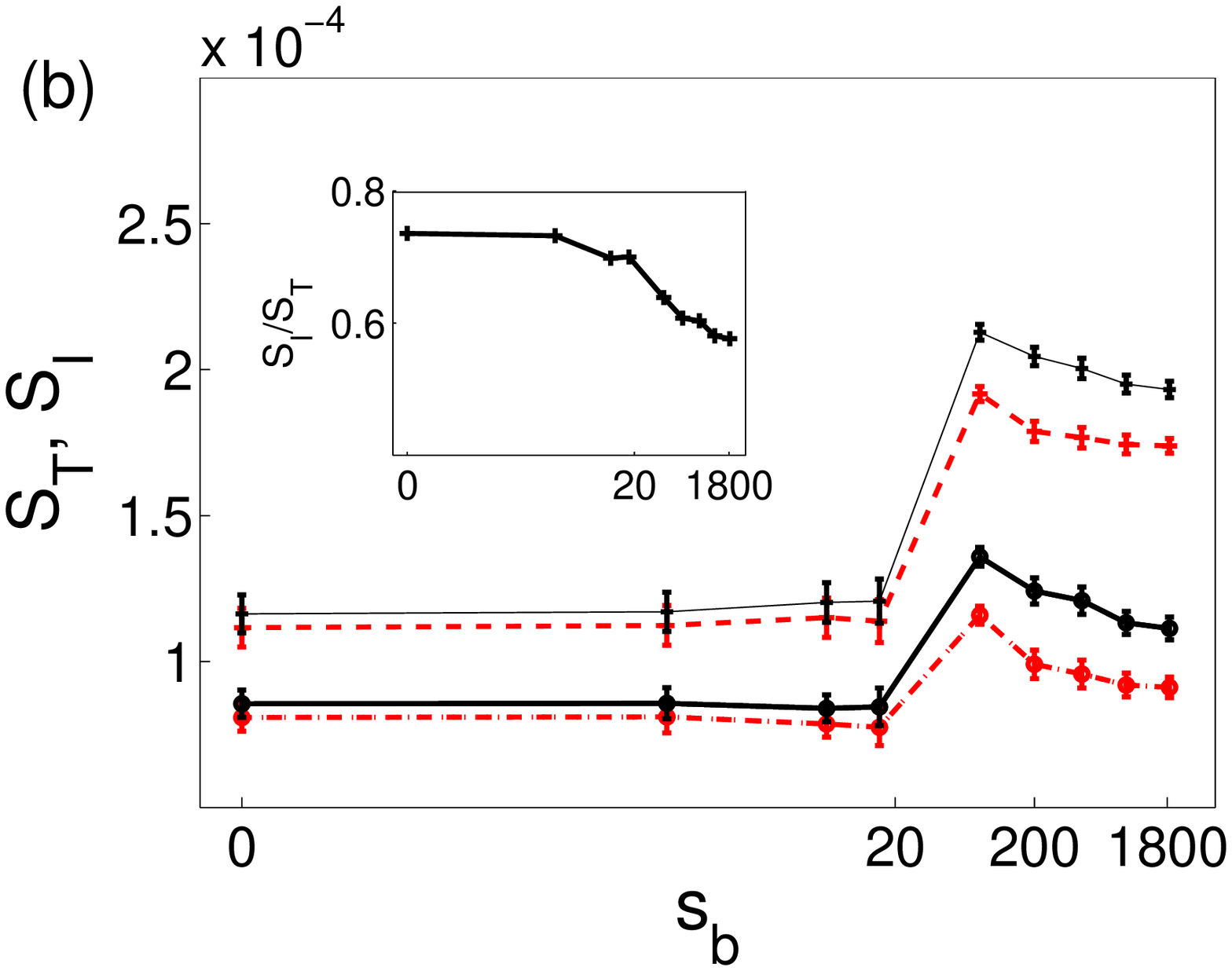}
\includegraphics[width=0.45\columnwidth]{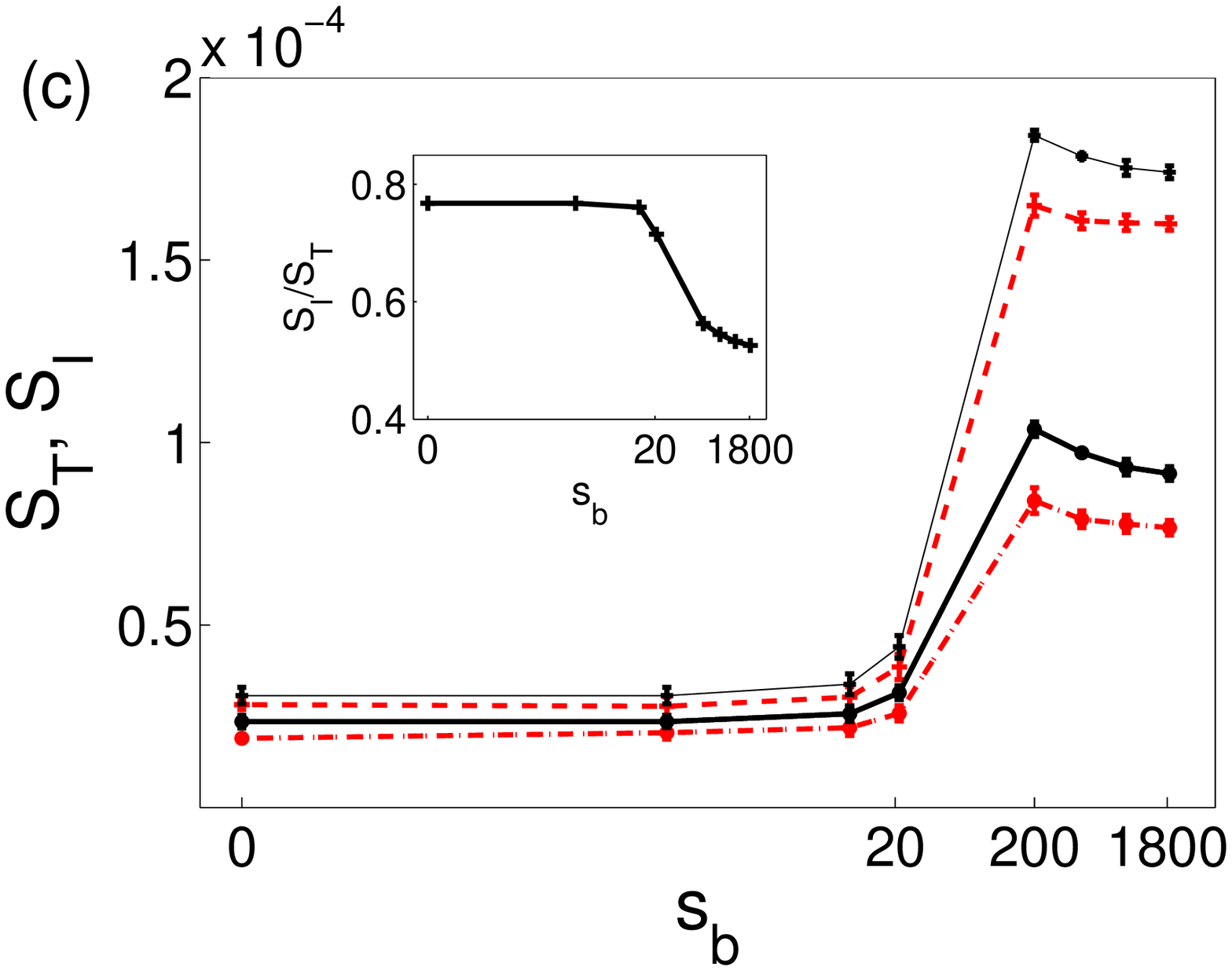}\includegraphics[width=0.45\columnwidth]{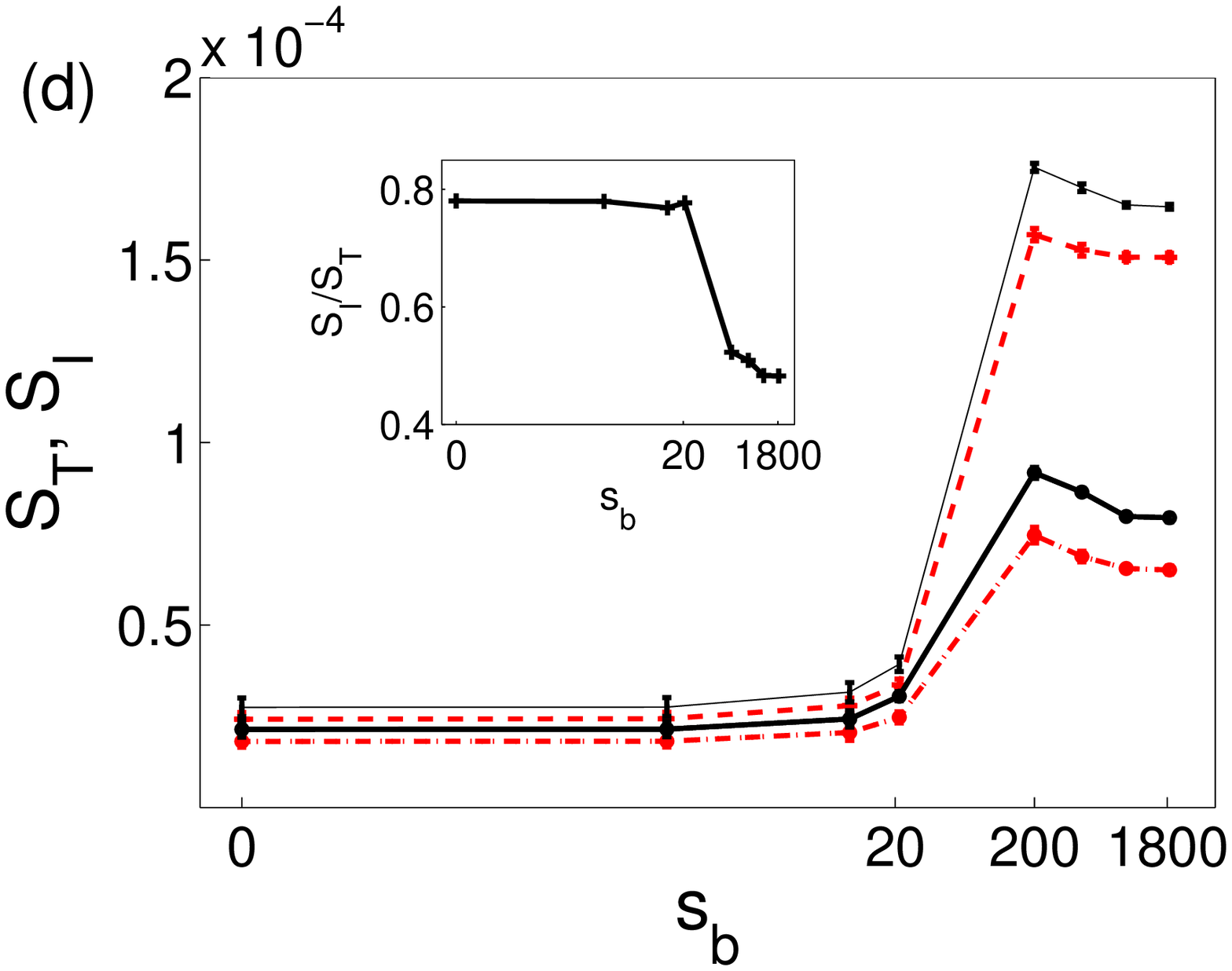}
 \end{center}
 \caption{\label{fig:2}
Estimates (dashed lines) derived from (\ref{eq:rhoest})
 and exact results (solid lines) for the total entropy per atom $S_{\rm T}$ (lines with crosses) and
  internal entropy per atom $S_{\rm I}$ (lines with circles) for different
  strengths of the scissors. 
Top row: Initial temperature $T=4$nK for the ramping speeds (from left
to right) 15.6 Hz/ms and 17.2 Hz/ms.
Bottom row: The same for $T=0$.
The curve for the total entropy per atom is shown
  together with the sampling error calculated numerically over 800
  individual stochastic realizations (see \ref{subsec:twa}).
The insets show the ratio
 $S_{\rm I}/S_{\rm T}$ (exact results) for each case. }
 \end{figure}

We evaluate the entropy per atom $S_{\rm I}$ in the
central four wells and and compare this
to the entropy per atom $S_{\rm T}$ in the full system.
The effect of different strengths of the scissors potential is shown in figure~\ref{fig:2}
by displaying the entropy in different cases at the end of the lattice ramping.
The simulation results of the barrier injection provide an example that demonstrates
how a single-particle entropy can be engineered in a nonequilibrium process by laser barriers.
Without scissors $S_{\rm I}<S_{\rm T}$, since in a combined optical lattice and
harmonic trap thermal phonon excitations and entropy are mostly concentrated
on the outermost sites. Decoupling those sites from the rest of the
system therefore reduces the effect of higher excitation modes.
For weak scissors ($s_b\lesssim 60$), the entropy does not notably increase and $S_{\rm I}$ remains lower
than the value of $S_{\rm T}$ without the potential barriers.
In the insets in figure~\ref{fig:2} we display the ratio $S_{\rm I}/S_{\rm T}$.
We find an increasing total entropy (but decreasing $S_{\rm I}/S_{\rm T}$)
for strong scissor potentials owing to excitations induced by injected barriers.
A detailed investigation of these excitations and resulting entropy waves, which is beyond the scope of
the present work, could in itself provide an interesting further study of entropy phenomena in a coupled
multi-mode BEC system.

\begin{figure}[t]
  \begin{center}
\includegraphics[width=0.45\columnwidth]{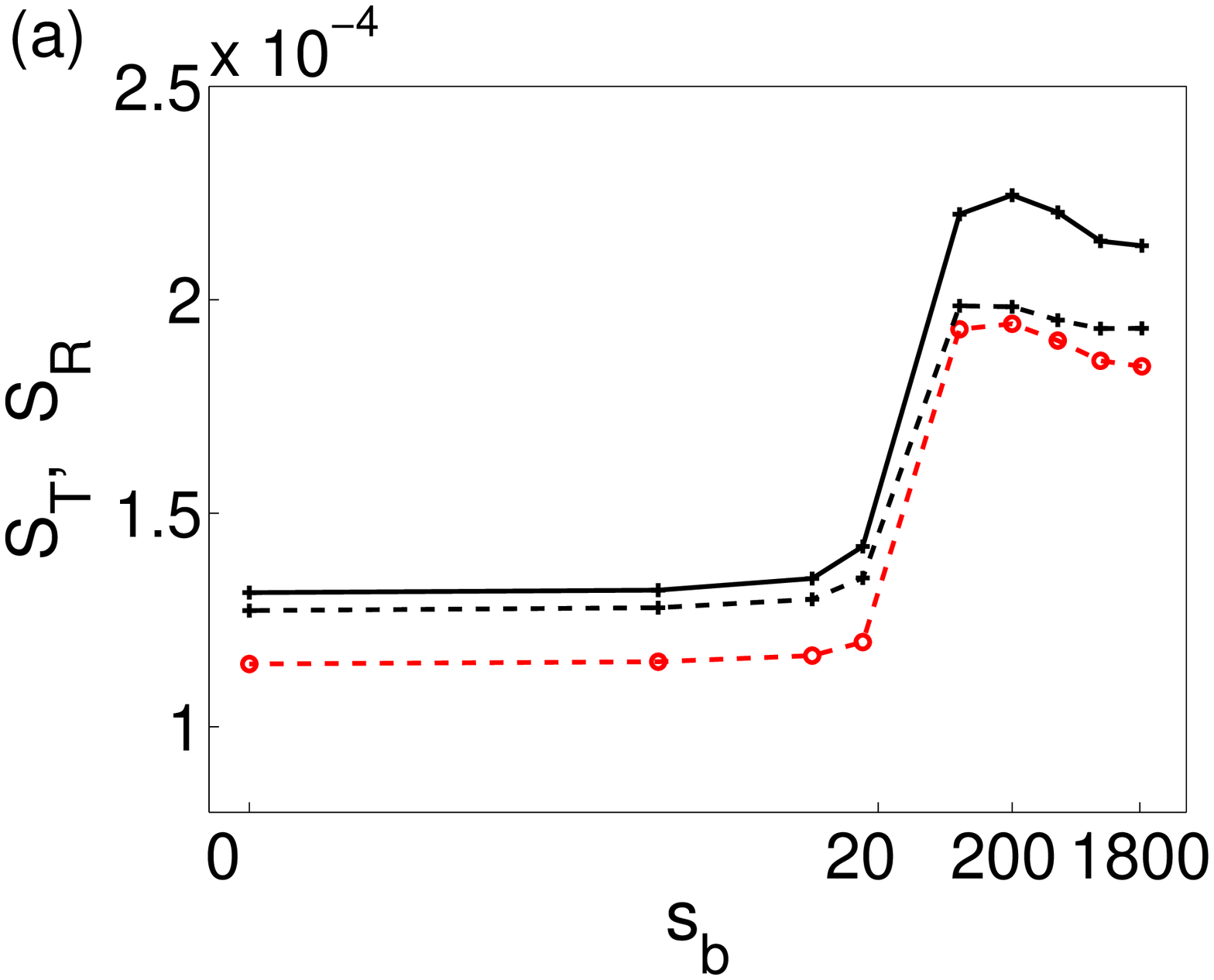}
\includegraphics[width=0.45\columnwidth]{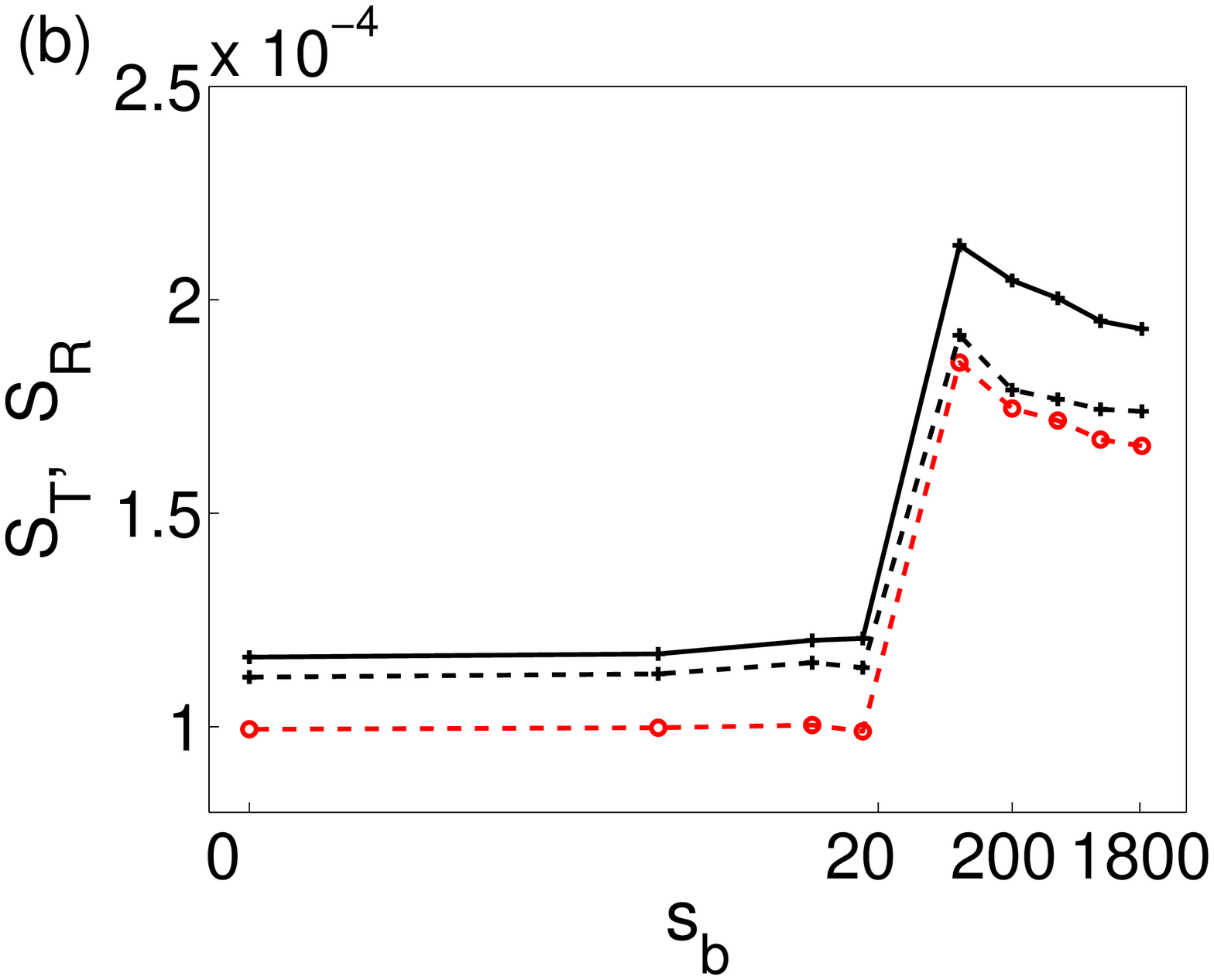}
  \end{center}
  \caption{\label{fig:som1}
Exact results for the total entropy $S_{\rm T}$ (solid lines) and its
estimates (dashed lines)
 for different  strengths of the barrier. The dashed black lines with
crosses show the estimate derived from our approximate expression of
the density matrix based on direct experimental observables (\ref{eq:rhoest}); the
red lines with circles show the estimate
$S_{\rm R}$ from equation (\ref{eq:renyi}) based on the R\'{e}nyi
entropies .
Initial temperature $T=4$nK  for the ramping speeds (from left to
right) $15.6$ Hz/ms and $17.2$ Hz/ms.}
\end{figure}

As it is not practical to measure the entropy by detecting all the vibrational
mode amplitudes, we propose an entropy estimate based on the lattice site occupation numbers and
the long-range coherence values.
In figure~\ref{fig:2} we compare the entropy
calculated from the $6\times 6$ density
matrix estimate $p_{\rm est}(\alpha,\beta)$ that is built by measuring separately only the
average well populations and the relative phases between the atoms in
the 6 different sites
\begin{equation}\label{eq:rhoest}
p_{\rm est}(\alpha,\beta) = \frac{\sqrt{\langle \hat{n}_\alpha \rangle  \langle \hat{n}_\beta \rangle}
  \langle \exp [i(\hat\phi_\alpha-\hat\phi_\beta)]\rangle}{\langle \hat{N}_{\rm T}\rangle},
\end{equation}
where $ \hat{n}_\alpha $ and $ \hat\phi_\alpha$ denote the atom number and phase operators in the site $\alpha$, respectively,
and $ \hat{N}_{\rm T}$ represents the total atom number in the six sites.
The good agreement between
the estimated entropy and the one based on the full basis state
representation can be explained by negligible atom number pair correlations between different
lattice sites  $\langle \hat n_\alpha \hat n_\beta \rangle\simeq \langle \hat n_\alpha \rangle
\langle \hat n_\beta \rangle$ and correlations between the phases and atom numbers
$\langle \sqrt{ \hat n_\alpha \hat n_\beta} \exp
[i(\hat \phi_\alpha-\hat \phi_\beta)]\rangle\simeq \langle\sqrt{ \hat n_\alpha  \hat n_\beta}
\rangle  \langle \exp [i(\hat \phi_\alpha-\hat \phi_\beta)]\rangle$. Similar correlations
between the atoms in vibrational states within {\em the same} site
only weakly affect the entropy. In addition, the loss of phase
coherence between the vibrational levels $k,l$ within the same site is
small $\langle \exp [i(\hat \phi_{\alpha k}-\hat \phi_{\alpha l})]\rangle \gtrsim 0.99$.
The entropy approximation (\ref{eq:rhoest}) is better at low temperatures owing to the weaker effect of thermal fluctuations
on intra-site correlations between the atoms in different vibrational states. Analogously, stronger
quantum fluctuations at stronger nonlinearities can lead to larger deviations from the exact result.
Well-known estimates of von Neumann entropy are based on combinations of R\'{e}nyi
entropies~\cite{ref:renyi} (see~\ref{sec:renyi}). It is therefore interesting to compare our estimate
to the R\'{e}nyi entropy estimates.
We show in figure~\ref{fig:som1} how our entropy estimate based on experimental observables
provides for this system a more accurate approximation of the entropy
than the estimate $S_R$  based on the
R\'{e}nyi entropies.

Injecting laser barriers affects the achievable spin squeezing
between the central sites after turning up of the lattice.
We define the relative atom number squeezing between the atoms in
sites $\alpha$ and $\beta$ by
\begin{equation}
\xi_{N, (\alpha,\beta)}^2=\left[\Delta \left(\hat{n}_\alpha
  -\hat{n}_\beta\right)\right]^2 \frac{\langle \hat{n}_\alpha \rangle + \langle \hat{n}_\beta\rangle}{4 \langle
  \hat{n}_\alpha\rangle \langle \hat{n}_\beta\rangle}\,,
\end{equation}
where $\Delta \left(\hat n_\alpha
-\hat n_\beta\right)$ denotes the relative atom number
fluctuations.
The spin squeezing of the atoms,
\begin{equation}
\xi_{S, (\alpha,\beta)}^2 \simeq {\xi_{N,
  (\alpha,\beta)}^2 \over  \langle \cos (\hat\phi_{\beta} -\hat\phi_{\alpha})\rangle^2} \,,
\end{equation}
not only depends on  $\xi_{N, (\alpha,\beta)}^2$, but also on the
relative phase coherence $\langle \cos (\hat\phi_{\beta} -\hat\phi_{\alpha}) \rangle$ between
the atoms. The proposed long-range phase coherence measurement scheme allows the detection of spin
squeezing also between non-nearest-neighbour sites and other spatially separated regions. In quantum-enhanced  metrology a high-precision quantum
interferometer can overcome the standard quantum limit of
classical interferometers, provided that $\xi_{S, (\alpha,\beta)}<1$
\cite{wineland, bouyer, holland,giovannetti,ref:gross}. The same
condition also implies quantum many-body entanglement in the
system~\cite{Sorensen:2001aa}.

\begin{figure}[t]
 \begin{center}
\includegraphics[width=0.45\columnwidth]{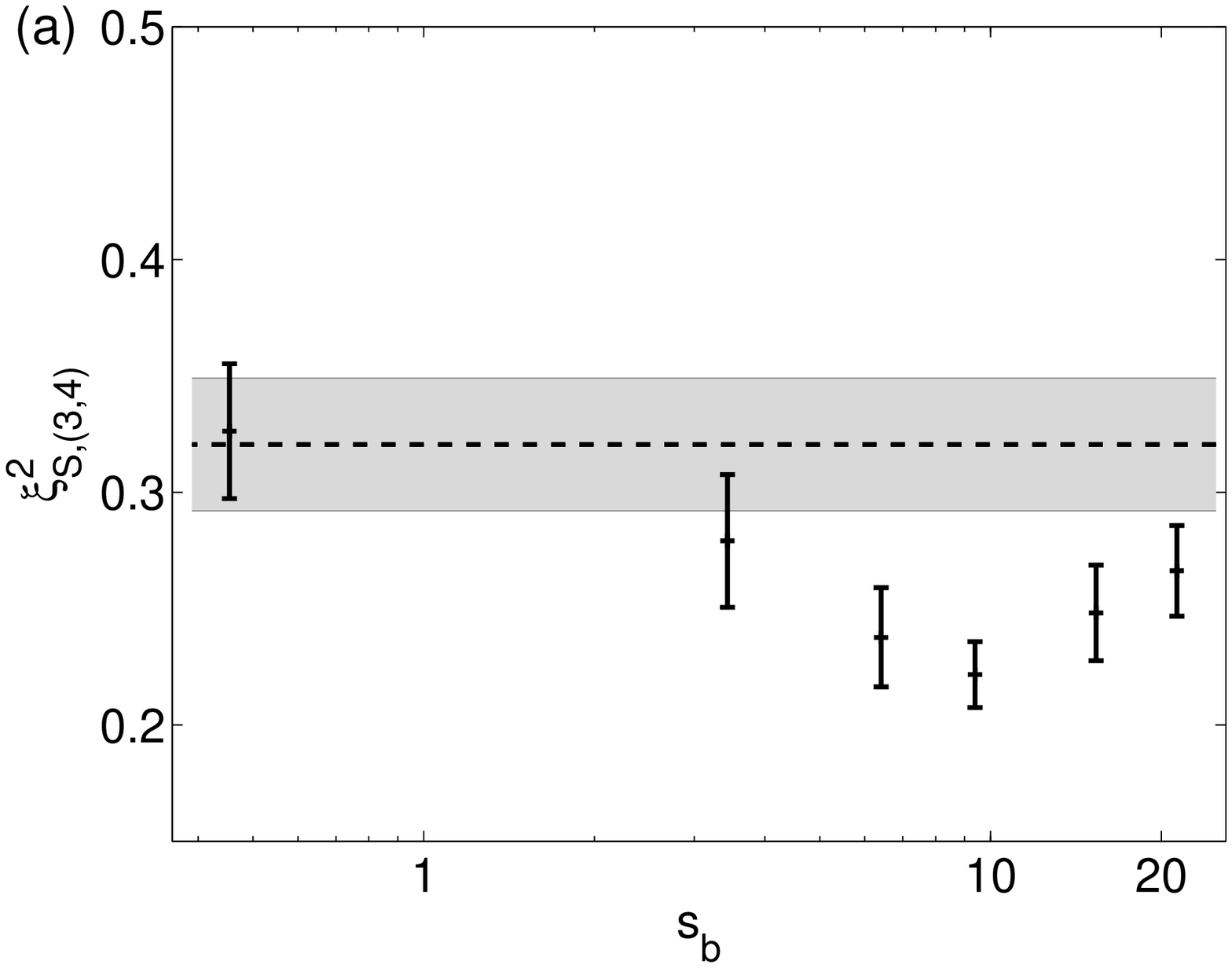}\includegraphics[width=0.45\columnwidth]{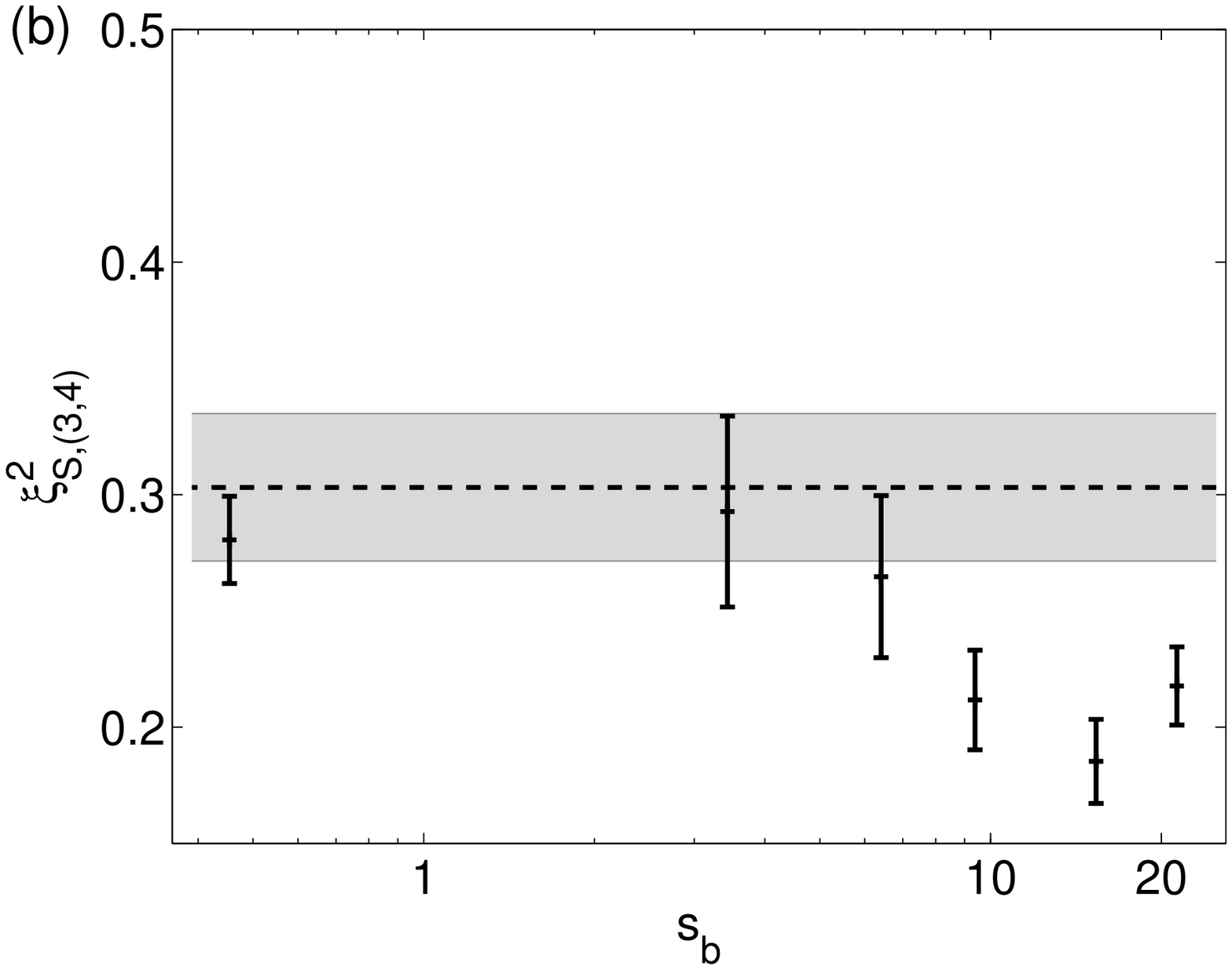}
\includegraphics[width=0.45\columnwidth]{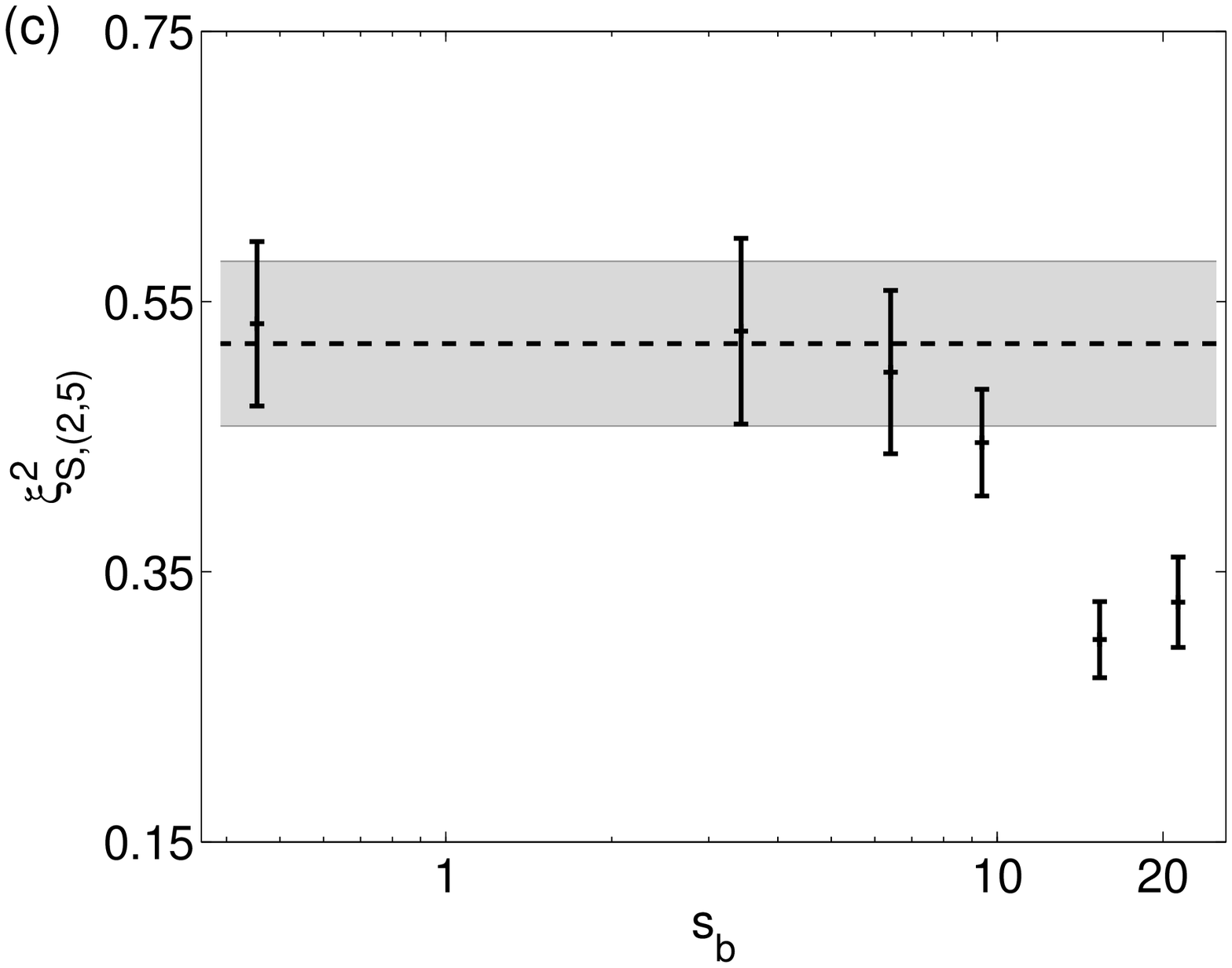}\includegraphics[width=0.45\columnwidth]{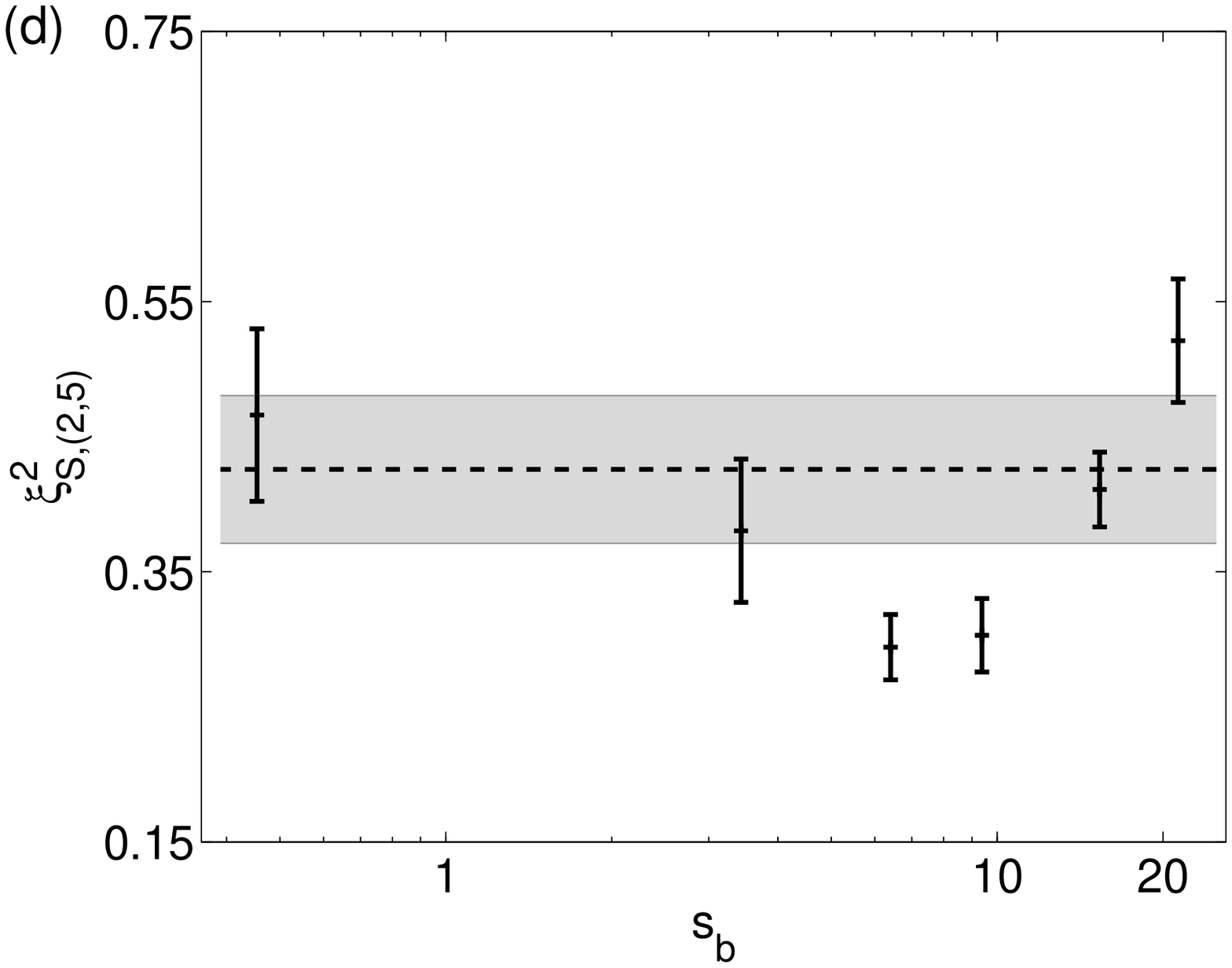}
 \end{center}
 \caption{\label{fig:chin}
   Top row: Spin squeezing $\xi_S^2$ between the atoms in the two
   central lattice sites (sites 3 and 4 in figure~\ref{fig:1}) for
   different strengths of the scissors potential, with the ramping
   speeds (from left to right) 15.6 Hz/ms and 17.2 Hz/ms.
   Bottom row: Spin squeezing between the atoms in
   non-nearest-neighbour sites (sites 2 and 5 in figure~\ref{fig:1}),
   with the ramping speeds (from left to right) 15.6 Hz/ms and
   17.2 Hz/ms.
   The dashed line shows the value obtained with no laser beam potential and the grey shaded area
   represents its uncertainty.
   The lattice is slowly ramped up to $72 E_R$ after the barriers are rapidly introduced in thermal equilibrium
 at the lattice height $24 E_R$ and $T = 4$nK.
 }
 \end{figure}

Results for the spin squeezing between the atoms in the two adjacent central
wells (sites 3 and 4 of figure~\ref{fig:1}) at the end of the ramping
for two different ramping speeds are shown in figure~\ref{fig:chin}(a)
and~(b) at the experimentally relevant initial temperature
$T=4$nK~\cite{ref:gross_pra}\footnote{Note that
  the actual experimental temperature in 3D trap may be higher than
  the one corresponding to experimental findings in 1D simulations.}.
At intermediate scissor strengths $s_b$, when the
nonadiabatic injection does not significantly perturb the system
[cf.~figure~\ref{fig:2}(a)], we find slightly stronger spin (as well
as the relative atom number) squeezing than in the system where no
laser potential was applied.
The spin squeezing between the atoms in non-nearest-neighbour sites
(sites 2 and 5 of figure~\ref{fig:1}) is shown in
figure~\ref{fig:chin}(c) and~(d).
The system exhibits spin squeezing and quantum many-body entanglement
between spatially separated regions specified by the distant
sites. Weak excitations of the system due to lattice and barrier
ramping affect the squeezing as can be seen from the differences between the two ramping speed
cases.  At stronger scissor strengths the barriers perturb the system,
resulting in notably stronger dynamics of the
squeezing.
We also find that in the region of improved spin squeezing the
relative phase coherence is not notably affected.
Our analysis provides a proof-of-principle demonstration that
adjusting the coupling between the inner and outermost sites could
potentially lead to technologies for improving atomic spin squeezing.

Spin squeezing flags the presence of entanglement in the system.
As an alternative signature of entanglement one can use entropic inequalities
in a bipartitite system~\cite{ref:horodecki,ref:alves1,ref:alves2}.
The purity ${\rm Tr}(\rho_{l(r)}^2)$ of the reduced density matrices of the left
(right) part of the system,
compared to the purity of the combined system ${\rm Tr}[(\rho_l
  \otimes \rho_r)^2]$, signals entanglement if ${\rm
  Tr}(\rho_{l(r)}^2) < {\rm Tr}[(\rho_l \otimes \rho_r)^2]$. Even
though we find spin squeezing in our system,
the purity based entanglement inequality is not
violated (see~\ref{sec:ent}). Remarkably, however, the purity estimates
derived from (\ref{eq:rhoest}) agree well (figure \ref{fig:som2}) with
the exact results (within $1\,\%$ for $s_b\lesssim 20$ and within $6\,\%$
for all values of $s_b$), thus potentially providing a direct
experimental detection method for the presence of entanglement.

In conclusion, we have shown how the single-particle entropy can be
measured and engineered in an optical lattice in the large filling
limit. The measurement can be achieved by matter-wave homodyne
detection of long-range phase coherence using a reference
condensate. The proposed scheme provides a good approximation of the
density matrix such that it might be used to detect entanglement via entropic inequalities.
We provided a proof-of-principle demonstration that controlling tunneling in selected
locations of the lattice may find quantum technological applications in manipulating
metrologically important spin squeezing.
Improvement in spin squeezing has practical implications since in the
experiments it is typically limited by thermal fluctuations~\cite{ref:gross}.
We may also envisage a procedure where the atomic spin squeezing is improved during
every injection cycle of the laser barriers that is followed by an adiabatic adjustment
of the trapping potential. Iterating the process could then potentially lead to progressively stronger
spin squeezing.

\section*{Appendix}

\appendix
\section{Stochastic phase-space method}\label{sec:twa}

\setcounter{section}{1}

\subsection{Dynamics}\label{subsec:twa}

In order to simulate the experimental conditions, quantum and thermal fluctuations are included via a
classical stochastic field description based on TWA
\cite{DRU93,Steel,SIN02,ref:isella06,ref:review,polkovnikov,martin10}.
TWA has provided a useful methodology for analyzing relative atom
number correlations between different lattice
sites~\cite{ref:gross_pra} and, e.g., between different solitons in
interferometric applications~\cite{brightsoliton}.
The dynamics is unraveled into stochastic trajectories where the initial state of the quantum
operator $\hat{\psi}(x,t=0)$ is represented by an ensemble of classical fields  $\psi_W(x,t=0)$,
sampled according to its Wigner distribution that is determined by the temperature,
the atom statistics and the multimode nature of the
system. The initial state $\hat{\psi}(x,t=0)$ is expanded in terms of the BEC ground state and the excited
states as
\beq\label{eq:initstate}
\hat{\psi}(x,t=0)=\psi_0 \hat{\alpha}_0 + \sum_{j>0} \left[u_j(x)\hat{
    \alpha}_j - v_j^*(x)\hat{\alpha}_j^\dagger\right].
\end{equation}
Here $\hat{\alpha}_j$ represent the quasiparticle annihilation operators and the
quasiparticle mode functions $u_j(x), v_j(x)$ are calculated by
solving self-consistently the coupled Hartree-Fock-Bogoliubov
equations \cite{ref:hfb1,ref:HFB} for the condensate and non-condensate
populations.
The quasiparticle operators are than replaced by stochastic complex variables $\alpha_j, \alpha_j^*$,
obtained by sampling the corresponding Wigner distribution. Each individual stochastic
realization is dynamically evolved according to the Gross-Pitaevskii equation
and represents a possible outcome of an individual experimental run. Ensemble averages
calculated from the TWA numerical results give a statistical description of the dynamics of
the system.

\subsection{Analysis of correlations}

The multi-mode nature of the system and phonon-phonon interactions
were shown to be important for the evaluation of the atom number
fluctuations \cite{ref:gross_pra}. Multi--mode effects are included by
projecting the stochastic field $\psi_W$ onto a mode function basis
formed by energy eigenfunctions of the lattice sites for a given number of
energy bands. The projection technique allows  to transform the
symmetrically-ordered expectation values of stochastic representations
of quantum operators in the Wigner distribution to normally-ordered
expectation values \cite{ref:gross_pra,ref:isella06,twabook,ISE05}. The projected
quantities are used in calculation of
the atom number statistics and the phase coherence of the system.
This also provides a model for the single particle entropy, evaluated
from the single-particle density matrix in the same mode function
basis.

We define an eigenmode basis for each site given by the mode functions
$\varphi_{\eta,j}(x)$ with $ \eta=1,\ldots, N_{w}, j=1,\ldots, N_{m}$. Here
the first index runs over all the sites, with $N_{w}=6$ in our case,
and the
second index runs over all the vibrational state mode functions in
individual sites.
We denote the stochastic amplitude for the atoms in the $j$th
vibrational mode of site $\eta$ as $a_{\eta,j}$ which can be
numerically obtained from the projection of the stochastic field
$\psi_W(x,t)$  as
\begin{equation}
a_{\eta,j}(t)=\bra \varphi_{\eta,j}|\psi_W(t)\ket = \int_{\eta^{\rm th} {\rm well}}
[\varphi_{\eta,j}(x)]^*\psi_W(x,t) dx\,.
\end{equation}
The macroscopic phase for the atoms in each site may be calculated by averaging over the
vibrational states
\begin{equation}\label{eq:phase}
\phi_\eta(t) \simeq \textrm{arg}\int_{\eta^{\rm th} {\rm well}} \left(
\sum_{j=1}^{N_m} a_{\eta,j}(t) \varphi_{\eta,j}(x)\right) \; dx\,.
\end{equation}
The relative phase coherence between the atoms in the sites $\eta$ and $\mu$ can then be obtained from
\begin{equation}
\Delta \phi_{\eta\mu}=\bra \cos \left(\phi_\eta-\phi_\mu\right) \ket_W\,,
\end{equation}
where the subscript $W$ denotes the Wigner expectation value over many realizations.
The projected amplitudes may be used to calculate the various normally
ordered quantum expectation values. The site populations read
\begin{equation}
\langle\hat n_\eta\rangle= \sum_j \langle \hat a_{\eta,j}^\dagger \hat
a_{\eta,j}\rangle=\sum_j \big[ \langle a_{\eta,j}^*  a_{\eta,j}\rangle_W-1/2
  \big],
\end{equation}
where the summation is over all the vibrational modes in the site $\eta$. The contribution
$-1/2$ in the last term is a result of the
symmetrical ordering of Wigner expectation values.
The on-site atom number fluctuations for site $\eta$ are similarly given by
\begin{eqnarray}\label{eq:onsite}
(\Delta n_\eta)^2 && =\sum_{i,k} [ \bra |a_{\eta,i}|^2 |a_{\eta,k}|^2\ket_W 
  \bra |a_{\eta,i}|^2\ket_W \bra |a_{\eta,k}|^2\ket_W \nonumber \\ 
&&-\delta_{i,k}/4]\,,
\end{eqnarray}
whereas the relative atom number fluctuations between two sites $\eta$ and $\mu$
are obtained from
\begin{eqnarray}\label{eq:deltan}
\Delta_{\eta\mu}^2 && \equiv  \left[\Delta \left(\hat{n}_\eta
  -\hat{n}_\mu\right)\right]^2 \\ \nonumber
  &&= \sum_{i,k}\big[\bra \left(|a_{\eta,i}|^2
  -|a_{\mu,i}|^2\right)\left(|a_{\eta,k}|^2-|a_{\mu,k}|^2\right)\ket_W \\ \nonumber
&&- \bra |a_{\eta,i}|^2-|a_{\mu,i}|^2\ket_W \bra |a_{\eta,k}|^2-|a_{\mu,k}|^2\ket_W-\frac{\delta_{i,k}}{2}\big]\,.
\end{eqnarray}

The single-particle density matrix $\hat{\rho}$ can be evaluated from the projected amplitudes
and it is given by the ensemble average of the stochastic realizations $\hat{\rho}_k$
where
\begin{equation}
\hat\rho_k=\sum_{\alpha j \beta l}p^{(k)}_{\alpha j,\beta l} |\varphi_{\alpha,j} \rangle\langle\varphi_{\beta,l}|\,.
\end{equation}
The matrix elements $p^{(k)}_{\alpha j,\beta l}$ are given by
\begin{equation}\label{eq:p}
p^{(k)}_{\alpha j,\beta l} = \frac{1}{\langle \hat N_{\rm T}\rangle }\left(a^*_{\alpha,j} a_{\beta,l}-\frac{\delta_{\alpha,\beta}\delta_{j,l}}{2}\right).
\end{equation}
Here $\langle \hat N_{\rm T}\rangle$ denotes the atom number in the studied six central sites.
The diagonal elements are determined by the mode
populations and the off-diagonal elements contain information on the
phase coherence.

\section{Entropy}\label{sec:renyi}

We first calculate the single-particle (generalized von
Neumann~\cite{ref:entropy}) entropy  $S$ of equation
(\ref{eq:entropy}) numerically in the projected single-particle basis
and then propose an estimate of the
entropy based only on direct experimental observables
\cite{ref:gross,ref:gross_pra}. After injecting the barriers the
entropy per atom $S_{\rm I}$ can be evaluated in the new reduced
lattice system of the four sentral sites,
\begin{equation}\label{eq:sint}
S_{\rm I} = \frac{\sum'_{\eta} \int_{\eta^{\rm th} {\rm well}} S(x) dx}{
  \sum'_\eta \bra \hat n_{\eta} \ket}\,,
\end{equation}
where the prime in the sum indicates a summation over the
wells that are inside the scissors and $S(x)$ the
entropy density.
We compare this with the entropy per atom $S_{\rm T}$ of the full system
\begin{equation}\label{eq:stot}
S_{\rm T} = \frac{\sum_{\eta} \int_{\eta^{\rm th} {\rm well}} S(x) dx}{
 \sum_\eta \bra \hat n_{\eta} \ket},
\end{equation}
where the summation is now over all the sites.

For comparison we also calculate an estimate of the von Neumann
entropy provided by a combination of the R\'{e}nyi entropies.
The R\'{e}nyi entropy of order $n$ is given by \cite{ref:renyi0}
\begin{equation}
S_n={1\over 1-n}\log(\textrm{Tr}(\hat{\rho}^n)), \quad n \ge 2\,.
\end{equation}
In the limit $n\rightarrow 1$, it coincides with the von Neumann
entropy.
The R\'{e}nyi entropy can be used to
approximate the von Neumann entropy~\cite{ref:renyi2}. Specifically,
an estimate $S_R$  of the von
Neumann entropy in terms of the R\'{e}nyi entropy can be given as 
\begin{equation}\label{eq:renyi}
S_R=\frac{1}{2}S_u+\frac{1}{2}\textrm{max}\left[S_{d23},S^d_{12}\right],
\end{equation}
where $S_u, S_{d23},S^d_{12}$ are appropriately defined functions of
$S_2$ and $S_3$ (see equation (32) in \cite{ref:renyi}).
We find that our entropy estimate that is based on experimental observables
provides a more accurate approximation of the entropy than the one based on the
R\'{e}nyi entropies, as shown in figure \ref{fig:som1}.

\section{Entanglement}\label{sec:ent}

\begin{figure}[t]
  \begin{center}
    \includegraphics[width=0.45\columnwidth]{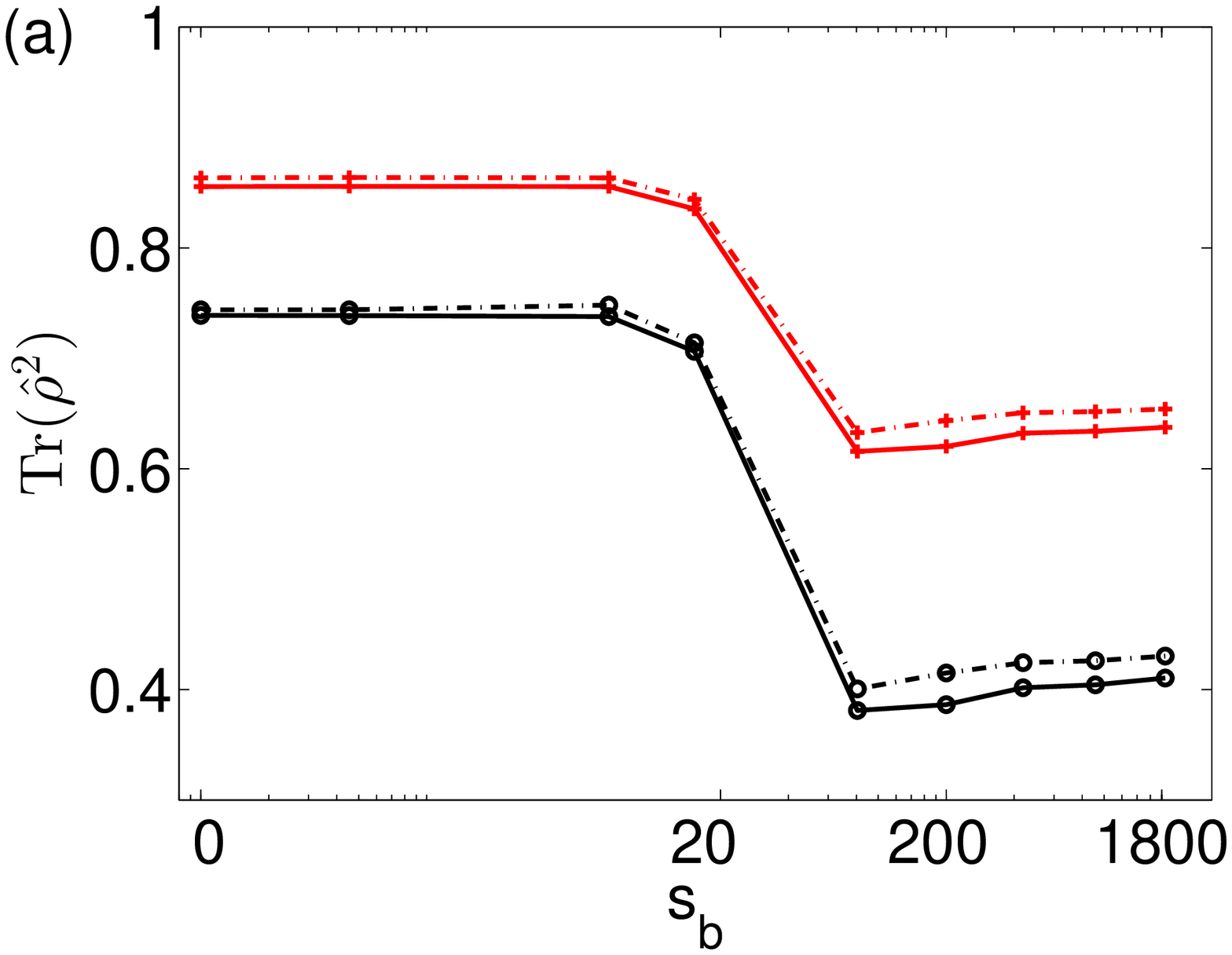}
       \includegraphics[width=0.45\columnwidth]{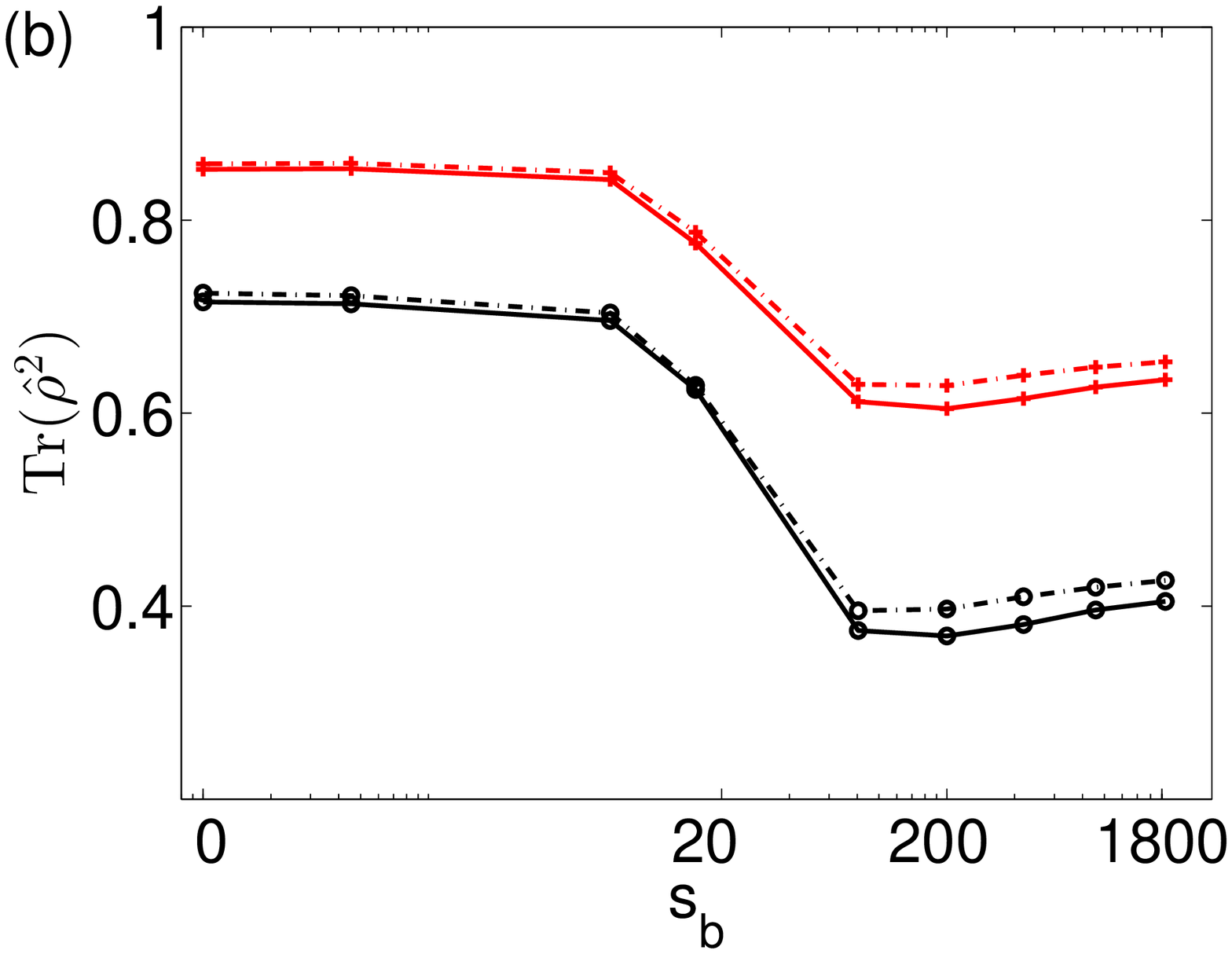}
  \end{center}
  \caption{\label{fig:som2}
Exact results (solid lines) and estimates (dashed lines) for the
purity  of system A (Tr$\hat{\rho}^2_A$, red lines with crosses) and
 of the combined bipartite system C (Tr$\hat{\rho}^2_C$, black lines with circles)
 for different  strengths of the barrier for the ramping speeds (from
 left to right) 15.6 Hz/ms and 17.2 Hz/ms. The dashed lines
show the estimates derived from our approximate expression of
the density matrix based on direct experimental observables (\ref{eq:rhoest}).
Initial temperature $T=4$nK. The relative sampling errors for the exact results are smaller than those
shown in figure \ref{fig:2}.
}
\end{figure}

Constructing the density matrix has useful applications, e.g., for
identifying bipartite entanglement
in the system~\cite{ref:horodecki,ref:alves2}. A simple test
 can be done
by comparing the purity of the full system described by $\hat{\rho}_C=\hat{\rho}_A\otimes
\hat{\rho}_B$  to that of a subsystem (described by the reduced density matrix $\hat{\rho}_{A(B)}$).
Separability of the two subsystems requires
Tr$\left(\hat{\rho}_{A(B)}^2\right)
\ge {\rm Tr}\left(\hat{\rho}_C^2\right)$.
Since there are only a few experimentally accessible tests for entanglement in many body systems, it is interesting
to compare the purities obtained from the proposed estimate of the
density matrix to the exact results. As an example, we define the left
half of the lattice system as subsystem A, the right half as subsystem B.
We calculate the purity of the reduced density matrix of the left--half subsystem A
and compare it to the purity of the combined bipartite system.
As shown in figure~\ref{fig:som2} this test does not detect entanglement, but remarkably the
results calculated from our density matrix estimate (\ref{eq:rhoest}) based on experimental observables provides a good approximation
of the exact result: The purities agree within $1\,\%$ for low barrier heights (below $s_b=20$) and within $6\,\%$ for all values of $s_b$.

\section*{Acknowledgements}
\begin{acknowledgements}
We acknowledge discussions with A.\ D.\ Martin and financial support by the Leverhulme Trust.
\end{acknowledgements}

\section*{References}
\bibliographystyle{unsrt}

\end{document}